# Surface superconductivity in the Weyl semimetal MoTe₂ detected by point contact spectroscopy

Yurii Naidyuk[1], Oksana Kvitnitskaya[1], Dmytro Bashlakov[1], Saicharan Aswartham[2], Igor Morozov[2,3], Ivan Chernyavskii [2,3], Günter Fuchs[2], Stefan-Lüdwig Drechsler[2], Ruben Hühne[2], Kornelius Nielsch[2], Bernd Büchner[2], and Dmitriy Efremov[2]

[1]*B. Verkin Institute for Low Temperature Physics and Engineering, NAS of Ukraine, 61103 Kharkiv, Ukraine*

[2]*Leibniz Institute for Solid State and Materials Research, IFW Dresden, Helmholtzstraße 20, D-01069 Dresden, Germany*

[3]*Lomonosov Moscow State University, Moscow, 119991, Russian Federation*

MoTe$_2$ is a Weyl semimetal, which exhibits unique non-saturating magnetoresistance and strongly reinforced superconductivity under pressure. Here, we demonstrate that a novel mesoscopic superconductivity at ambient pressure arises on the surface of MoTe$_2$ with a critical temperature up to 5 K significantly exceeding the bulk $T_c$ = 0.1 K. We measured the derivatives of *I–V* curves for hetero-contacts of MoTe$_2$ with Ag or Cu, homo-contacts of MoTe$_2$ as well as "soft" point contacts (PCs). Large number of these hetero-contacts exhibit a *dV/dI* dependence, which is characteristic for Andreev reflection. It allows us to determine the superconducting gap $\Delta$. The average gap values are 2$\Delta$=1.30±0.15 meV with a 2$\Delta/k_B T_c$ ratio of 3.7± 0.4, which slightly exceeds the standard BCS value of 3.52. Furthermore, the temperature dependence of the gap follows a BCS-like behavior, which points to a nodeless superconducting order parameter with some strong-coupling renormalization. Remarkably, the observation of a "gapless-like" single minimum in the *dV/dI* of "soft" PCs may indicate a topological superconducting state of the MoTe$_2$ surface as these contacts probe mainly the interface and avoid additional pressure effect. Therefore, MoTe$_2$ might be a suitable material to study new forms of topological superconductivity.

# Introduction

In spite of the fact that intensive studies on the transition metal dichalcogenides (TMD) started more than fifty years ago (see Wilson & Yoffe [1] for a comprehensive review with references therein), these materials still evoke surprises due to their interesting structural chemistry and unusual electronic properties. Recently, an enormous interest in layered TMD arose after the discovery of huge non-saturating magnetoresistances in $WTe_2$ [2] and $MoTe_2$ [3], pressure induced superconductivity [3, 4] and predicted Weyl semimetal states [5]. As to Weyl semimetals, they show a peculiar band structure both in the bulk and on the surface, which is responsible for their outstanding properties and might have promising prospects, e.g., for low dissipation quantum electronics and spintronics [6].

TMD crystallize in different crystallographic structures denoted as 2H-, $T_d$-, 1T-, and 1T`-type lattices. The 2H- and 1T-type compounds are semiconducting, whereas the $T_d$- and 1T`-type compounds show a semimetallic behavior. Among dozens of known TMD, the greatest attention is paid nowadays to $WTe_2$ and $MoTe_2$. Molybdenum ditelluride ($MoTe_2$) is one of the TMDs, which can be grown either in the semiconducting 2H or the semimetallic 1T` form. The latter phase of $MoTe_2$ is stable at room temperature, but transforms to the $T_d$ structure by cooling below 240 K [7]. It turned out that $MoTe_2$ in the $T_d$ structure is a Weyl semimetal of type II [8]. This was confirmed by subsequent extensive band structure calculations, ARPES and quasiparticle scattering measurements (see, e.g. Ref. [9, 10] for very recent publications and Refs. therein). However, there is a lack of transport measurements that can probe topological surface states directly.

It is intriguing also that $MoTe_2$ undergoes transition to the superconducting (SC) state with the critical temperature $T_c = 0.1$ K at ambient pressure, which increases up to 8.2 K under a pressure of 11.7 GPa [3]. The results immediately lead to the question, if the superconductivity involve also the edge states, especially in a mesoscopic geometry. The topologically protected edge states of Weyl semimetals may become superconducting due to the proximity effect to the bulk states and show nontrivial physical properties. This would open a new platform for the investigation of exotic topological superconductivity, which may have potential applications, e.g., in quantum computation [11, 12].

In this article, we show the emergence of unique superconducting properties at mesoscopic point contacts (PCs) between high quality single crystals of the Weyl semimetal $MoTe_2$ and conventional normal metals. We investigated both "hard" PCs prepared with Ag (or Cu) tips and "soft" type PCs [13], which were prepared by dripping of a small drop of silver paint onto the cleaved $MoTe_2$ surface to avoid pressure effect. In the former case the critical temperature rises at the interface up to 5 K, while in the latter case up to 4 K. Although in the case of the "hard" PC one might assign the strong increase of the SC critical temperature to the local pressure, almost the same increase of the critical temperature for the "soft" contacts, unambiguously points to pure interface effect. Applying PC Andreev-reflection (AR) spectroscopy [13, 14, 15, 16] and Yanson PC spectroscopy [15], we try to determine all characteristics of the observed SC state.

# Results

**Point contact spectroscopy with "hard" tips**. Figures 1 and 2 show the *dV/dI* curves for typical $MoTe_2$–Ag hetero-contact (hereinafter – "hard" contact) in dependence on their temperature and in magnetic field. About a dozen of "hard" contacts with either Ag or Cu (of total 20 "hard" contacts) show similar a *dV/dI* dependence with a characteristic AR structure, i.e., a double-minimum around zero-bias.

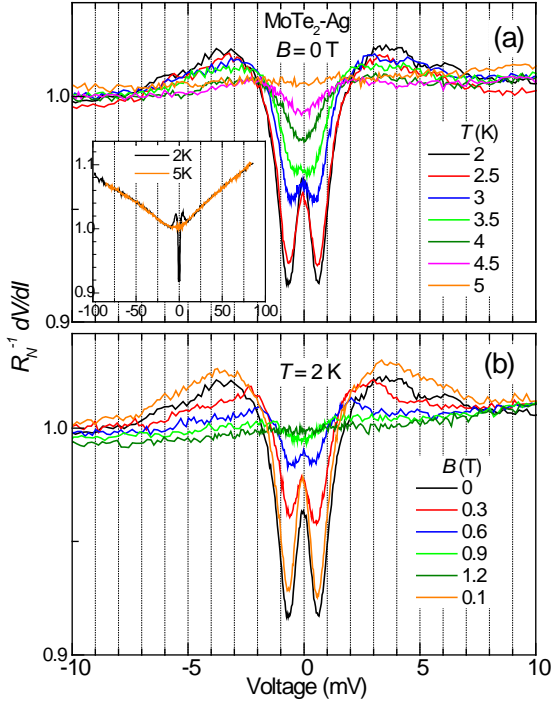 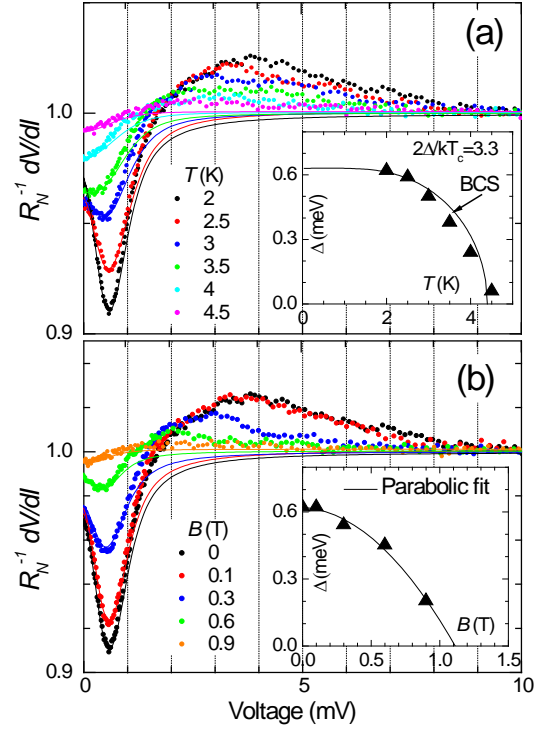

**Figure 1.** *dV/dI* spectra for a MoTe$_2$–Ag PC. Temperature (a) and magnetic field (b) dependence of *dV/dI* for a PC with a normal state resistance of 3 Ω. The inset shows two *dV/dI* (below and above $T_c$) at larger bias.

**Figure 2.** Fit of the normalized *dV/dI* spectra from Fig.1. Fit (thin lines) of *dV/dI* (symbols) for different temperatures at **B** = 0 (a) and magnetic field at *T* = 2 K (b). Insets show the temperature and magnetic field dependence of the SC gap from the fit.

With increase of temperature or magnetic field, the magnitude of the double-minimum gradually decreases and this structure vanishes at a particular temperature or magnetic field, which confirms its SC nature. The minima in *dV/dI* are located near the expected SC gap values [14]. The gap size as well as their temperature and magnetic field dependence can be obtained from the well-known fitting procedure [13, 16]. The calculations (i.e., the fit of *dV/dI*) were performed using a single gap Blonder-Tinkham-Klapwjik (BTK) theory [14]. The fit matches perfectly the region of the expected gap (minimum) position (see Figures 2(a) and (b)), but shows deviations for values above 2 mV, where broad side maxima develop. The latter features are most probably of non-Andreev nature and apparently connected with the destruction of superconductivity in the PC due to the high current density originating from the increasing bias voltage [16, 17]. The results of the calculations are present in Figures 2(a) and (b), where the *dV/dI* characteristics (normalized to the normal state *dV/dI*) and their fitting are shown for different temperatures and magnetic fields. Additionally, the temperature and magnetic field dependence of the extracted SC gap is displayed in the inset. The temperature dependence of the SC gap (Figure 2(a), inset) is close to the anticipated behavior for BCS superconductors, whereas the magnetic field dependence of SC gap (Figures 2(b), inset) follows $\Delta = \Delta_0\sqrt{1 - H/H_{c2}}$ [1] expected for type-II superconductors [19]. The critical temperature $T_c$ for this PC evaluated from the BCS fit is about 4.4 K and the critical field is about 1.1 T, as shown in the insets of Figure 2.

As the $T_c$ of bulk MoTe$_2$ is below 0.3 K[2] and increases up to 8.2 K under the pressure of 11.7 GPa [3], the obvious explanation for the enhanced $T_c$ in "hard" PC is related to the pressure effect. It might take place during the formation of PC by attaching the Ag or Cu electrode mechanically onto the MoTe$_2$ surface. To clarify this issue, we further investigated "soft" PCs.

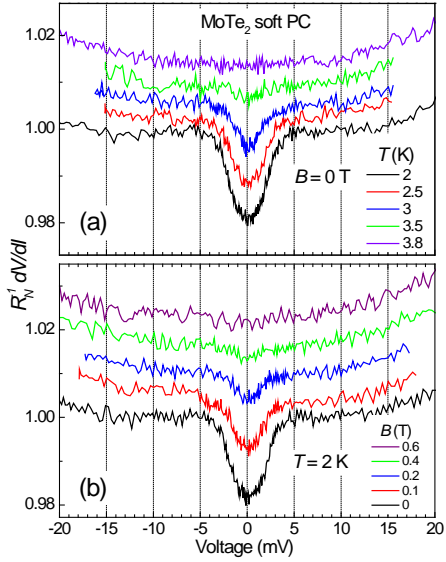 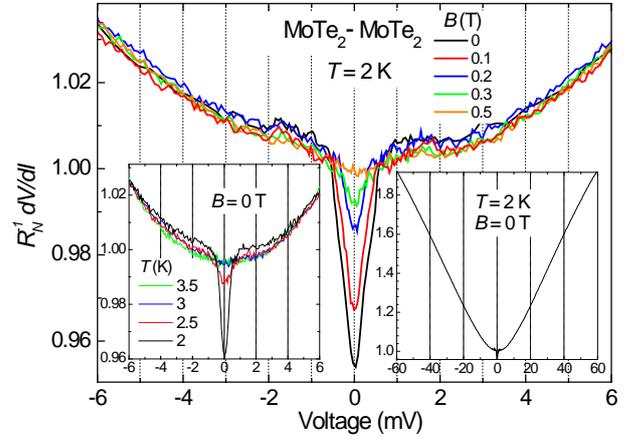

**Figure 3.** $dV/dI$ spectra for a "soft" MoTe$_2$ PC. Temperature (a) and magnetic field (b) dependence of the spectra for a PC with a normal state resistance of $R_N=0.7$ Ω. The curves are offset vertically for clarity.

**Figure 4.** $dV/dI$ spectra for a MoTe$_2$ homo-contact. Magnetic field (main panel) and temperature (left inset) dependence of the spectra for a homo-contact with a normal state resistance of $R_N=3.3$ Ω. The right inset shows $dV/dI$ for a larger bias.

**Point contact spectroscopy using "soft" and homo-contacts.** We prepared in total 10 so-called "soft" PC by dripping of a small drop of silver paint onto the cleaved MoTe$_2$ surface to avoid any pressure effect. Nevertheless, we again observed an increase of $T_c$ up to 4 K for these PCs (see Figure 3), although the intensity of the zero-bias minimum is smaller in this case. However, we did not observe $dV/dI$ characteristics with a typical AR double minimum as in the case of "hard" contacts. All measured "soft" PCs demonstrate a single approximately "V" shaped minimum, with a depth in the order of 1%, which is about one order of magnitude smaller than that of "hard" contacts (see Figure 1). The reason might arise from the fact that the "soft" PCs consist of an array of tiny contacts between microscopical silver grains (having a size of 2–10 μm [13]) and the MoTe$_2$ surface, therefore, individual PCs can be smaller compared to mechanical PCs. In this case, "soft" PCs examine indeed the surface properties.

Additionally, we also measured $dV/dI$ curves for 6 homo-contacts of MoTe$_2$ (see Figure 4), where a zero-bias minimum was observed in the $dV/dI$ characteristics, which vanishes close to a temperature of 3–4 K and in a magnetic field below 1 T. At the same time, no Josephson critical current was detected in the case of homo-contacts, so far, probably due to mesoscopic size of the SC region and a specific topological SC state at the interface.

---

[1] For some other PCs, the fit of the $dV/dI$ curves measured in magnetic field results in a weaker suppression of the SC gap value with field, although the magnetic field strongly suppresses the $dV/dI$ intensity (similar like in Fig. 1(b)). We have observed a similar behavior already for "soft" FeSe PCs [18].

[2] According to Rhodes et al. [20], $T_c$ in MoTe$_2$ depends on sample quality, increasing as the residual resistive ratio increases, suggesting that structural disorder suppresses $T_c$.

**Critical temperature and magnetic field and low temperature SC gap.** Figure 5 shows the distribution of the critical temperatures and the critical magnetic fields for the measured PCs. In contrast to the relatively narrow distribution of the critical temperature (Figure 5(a)) and the SC gap values (Figure 6(a)), the critical magnetic fields show a large scatter between 0.3 T and 2 T for these PCs. Within the distribution, the "soft" PCs have only a small scattering of the critical field, which concentrates near 0.4 T, while the "hard" PCs demonstrate higher values of the critical magnetic field. This again indicates the difference in the SC state of the "soft" and "hard" PCs.
Figure 6 shows the SC gap $\Delta$ and the ratio $2\Delta/k_BT_c$ extracted for all PCs with an AR structure using the BTK fit procedure. As a result, the $2\Delta/k_BT_c$ ratio is found to be on the average 3.7 +/- 0.4, i.e, slightly exceeds the BCS value of 3.52, which points to moderate strong coupling corrections due to the presence of electron-boson coupling. This $2\Delta/k_BT_c$ may also exclude node lines, which results in significantly higher ratio (two-dimensional $d$-wave leads to 4.2 [21]).

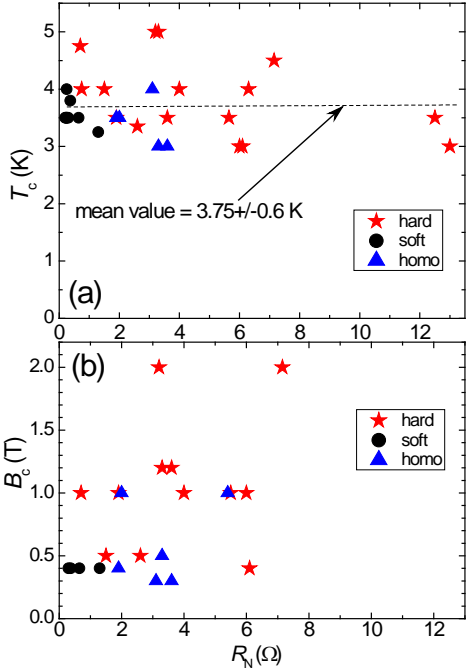
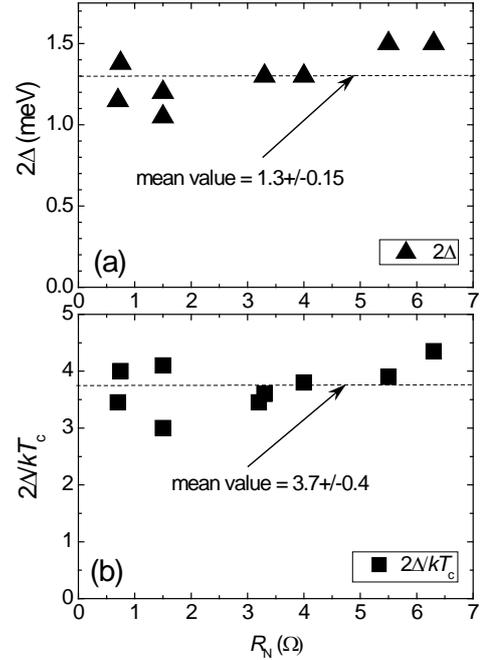

**Figure 5.** $T_c$ and $B_c$ distribution in MoTe$_2$ "hard", "soft" and homo-contacts. Critical temperatures (a) and critical magnetic fields (b) in dependence of the contact resistance for all measured PCs.

**Figure 6.** $2\Delta$ and $2\Delta/k_BT_c$ in MoTe$_2$ "hard" contacts. Calculated SC gap (a) and reduced gap ratios (b) for PC "hard" contacts, which demonstrate an AR double-minimum structure.

**Yanson PC spectroscopy in the normal state.** To investigate the origin of the cooper pairing, we have carried out Yanson PC spectroscopy measurements [15]. It is based on the fact that in the normal state the second derivative $d^2V/dI^2$ of the $I-V$ characteristic of ballistic PC is proportional to the electron-phonon interaction (EPI) function $\alpha^2F(\omega)$, i.e., $d^2V/dI^2 \sim \alpha^2F(\omega)$. The $d^2V/dI^2$ for the PC with the most pronounced maxima is shown on Figure 7. It is characterizes by a broad maximum between 15 and 20 mV, which can be attributed to the EPI. Sharp peaks close to zero-bias are due to residual superconductivity in the PC, which is not fully suppressed by the field of 0.4 T. Such "zero-bias" peaks are absent for PCs measured above $T_c$ (see inset on the bottom of Figure 7). Arrows below the $d^2V/dI^2$ spectrum indicate the position of the main maxima in the EPI spectrum of Ag [22] (which is used as a counter-electrode) and pure Mo [23]. As can be seen, the main phonon modes of the Ag tip do not give a discernible contribution to $d^2V/dI^2$ behavior; therefore, all features in the spectrum are connected to MoTe$_2$.

A recent calculation of the phonon spectrum for MoTe$_2$ [24] is shown at the bottom of Figure 7. The broad maximum in the $d^2V/dI^2$ spectrum between 15 and 20 mV corresponds well to the main peak in the phonon spectrum around 18mV, whereas the phonon maxima around 30 mV is not visible in the PC spectrum as well as the maxima below 10 mV. The reason for the latter might be weak interaction of electrons with these low energy phonon modes, while phonon features at higher energy are usually smeared out continuously in the $d^2V/dI^2$ spectra due to the shortening of the inelastic mean free path of electrons with increasing energy and corresponding rise of the scattering rates. All these features point to the fact that the maximum in the $d^2V/dI^2$ spectrum is indeed caused by the EPI in MoTe$_2$.

Thus, the PC EPI spectrum for MoTe$_2$ corresponds, in general, to the calculated phonon spectrum, but it is softer than that of clean Mo [23]. This is in accordance with the fact that the $T_c$ of Mo (which is slightly below 1 K according to [25]) is lower than the $T_c$ observed here for MoTe$_2$ PCs. One should note that the Raman spectra of MoTe$_2$ [3] display two main maxima with energies of 20 meV and 32 meV [3], respectively, which are also above the $d^2V/dI^2$ maximum. Thus, it seems that the EPI in MoTe$_2$ is modified in the PC or at the interface as compared to the bulk MoTe$_2$. This modification of the EPI (i.e., a softening of the EPI spectrum) might be one of the reasons for the observed enhanced $T_c$, but a more detailed study is needed.

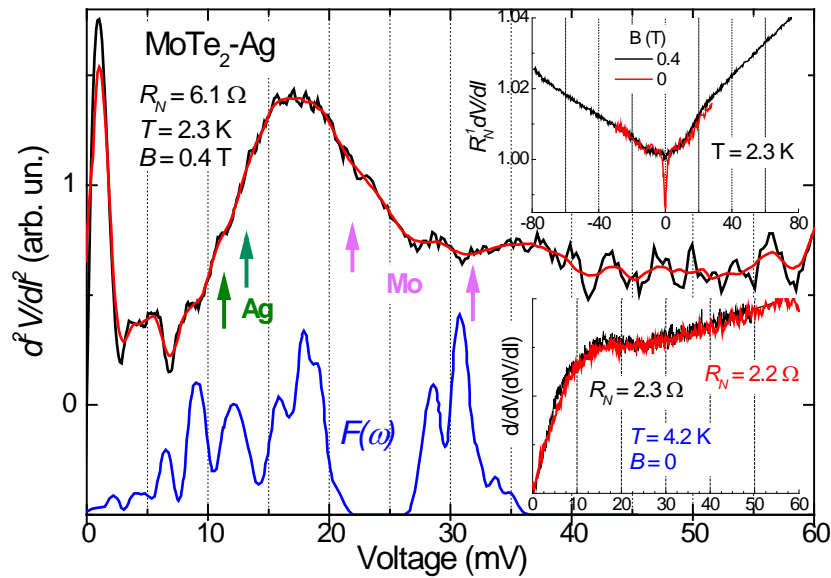

**Figure 7.** Yanson PC spectroscopy of MoTe$_2$. $d^2V/dI^2$ spectrum (black curve) for MoTe$_2$–Ag PC averaged for two polarities (i.e., current direction). The flowing red curve is a 10 point smoothing of the black one. Arrows show the position of the main maxima in the EPI spectrum of Ag [22] and Mo [23]. The bottom (blue) curve is the calculated phonon density of states $F(\omega)$ from Ref. [24]. The bottom inset shows the calculated derivative $d(dV/dI)/dV \sim d^2V/dI^2$ for two other MoTe$_2$–Ag PCs, that almost coincide and show a broader maximum between 10 and 20 mV. The upper inset shows $dV/dI$ curves for the same PC at zero field with a sharp zero-bias SC minimum (solid red curve) and at 0.4 T (dash black curve), where the zero-bias SC minimum is almost suppressed.

---

[3] It is curious that the Raman scattering data [3] almost coincide with the position of the main phonon peaks in the EPI spectrum of pure Mo at 22 and 32 meV [23], i.e., in spite of the different crystal structure of MoTe$_2$ and Mo the main phonon modes have similar energy. Maybe the reason is that, as mentioned in [7],"The Mo–Mo-distance of 2.89Å for β-MoTe$_2$ is only 0.19 Å larger than in the pure molybdenum crystals".

# Discussion and conclusion

We found a zero-bias minimum structure in the *dV/dI* spectra of PCs created on the surface of MoTe$_2$ single crystal. The SC origin of this structure is confirmed by the suppression of the *dV/dI* feature by temperature and magnetic field. The critical temperature derived from this observation is in the range of 3–5 K, which is an order of magnitude larger than the values for bulk MoTe$_2$ samples [3, 20]. It testifies that the superconductivity appears in the PC region. The observation of a similar SC state in the case of "soft" PC questioned the pressure effect as origin for the enhanced $T_c$ at least in the latter PCs. Therefore, we assume that the superconductivity emerges at the surface and it is probed by the surface sensitive PC measurements.

As shows the analysis of the observed AR structure for "hard" contacts (i.e., the zero-bias double minimum), the evaluated SC gap follows a BCS like behavior as found for common superconductors. The temperature dependence of the gap function obtained from the analysis of the Andreev reflection spectra can be fitted with a BCS curve. The averaged reduced gap value $2\Delta/k_BT_c = 3.7 +/- 0.4$ is close to the BCS value of 3.52. In a recent paper [26], the SC gap in MoTe$_2$ was evaluated using magnetic penetration depth measurements under pressure. For such pressure induced superconductivity a two gap model was proposed with $2\Delta_1/k_BT_c = 1.5$ and $2\Delta_2/k_BT_c = 4.6$ at a pressure of 0.45 GPa with a $T_c = 1.5$ K. Taking into account that the weighting factor of the large gap was estimated in Ref. [26] to be 0.87, the averaged $2\Delta/k_BT_c$ value will be 4.2, which is closer to our data shown in Figure 6(b).

Regarding the "soft" PCs, the disturbance of the surface and connected pressure effects are expected to be negligible. In the case of "soft" PCs, they probe to a greater extent the surface properties, as discussed above. However, in contrast to the "hard" contacts, we did not observe *dV/dI* spectra with the characteristic Andreev reflection double minima. All measured "soft" PCs demonstrate a single approximately "V" shaped minimum. Taking into account that topological superconductors possess a full pairing gap in the bulk and gapless surface states, the observation of "gapless-like" single minimum in the *dV/dI* of "soft" PCs may indicate the topologically governed superconductivity of MoTe$_2$ surface states.

We assume two possible scenarios for the origin of the surface superconductivity. The first scenario is based on the observation that the Weyl semimetals possess protected surfaces states. Indeed, let us estimate of the effective coupling constant in the framework of the BCS theory $T_c = 1.13\omega_D \exp(-1/\lambda_{eff})$. Here we suppose that the effective $\lambda_{eff}$ is the renormalized by the coulomb repulsion. Using the characteristic energy $\omega_D$ of 20 meV of the bosonic spectral function shown in Figure 7, we get the value of the coupling constant $\lambda_{eff} \sim 0.15$ in the bulk ($T_c \sim 0.3$K [3]) and $\lambda_{eff} \sim 0.25$ at the interface ($T_c \sim 5$K). This increase of $\lambda_{eff}$ by $\sim 60\%$ may arise from the increased density of states close to the surface due to topologically protected surface states in Weyl semimetals, which were predicted in the DFT calculations and observed in ARPES experiments [27]. Therefore, the mechanism should be universal for Weyl semimetals. Let's note, similar enhancement of the superconductivity in PCs was recently observed in type I Weyl semimetal TaAs [28, 29].

The second possible scenario is based on the observation that the low temperature $T_d$ crystal structure of MoTe$_2$ undergoes transition back to the room temperature 1T` phase under pressure [3, 30] with a strong increasing of $T_c$. It is possible that silver paint (some kind of glue) fixes the 1T` room temperature phase at the interface by cooling of "soft" PC to the low temperature. Thus, we have "high $T_c$" structure at the surface for "soft" PC in proximity with "low $T_c$" structure in the bulk. In this case, superconductivity of "soft" PCs may be under impact by nontrivial topology of the bulk $T_d$ phase. As to the SC states in the "hard" PCs, they have likely pressure induced origin within this scenario.

Another important question whether the superconductivity is topological. As shown by Rhodes et al. [20], SC state in MoTe$_2$ is very sensitive to the sample quality and $T_c$ onset stretches up to 0.7K for high quality samples. The sensitivity to the disorder gives the hope of dominating triplet component. But, this issue requires further study.

Summarizing, we have discovered a tremendously enhanced superconductivity in the potential Weyl semimetal MoTe$_2$ probed by surface sensitive PC measurements. Although the temperature dependence of the extracted gap function follows the BCS behavior in the case of "hard" PCs, the different "gapless-like" $dV/dI$ characteristics for "soft" PCs points to a non-trivial SC state of MoTe$_2$ surface, apparently influenced by topological Weyl states of the bulk. Albeit for elucidation of the microscopic nature of the superconductivity, as our discussion shows above, the role of the surface states needs further examination both experimentally and theoretically. Therefore, MoTe$_2$ is an attractive material to study new forms of superconductivity in topological environment.

## Methods.

**Synthesis.** Bulk single crystals of 1T´ MoTe$_2$ were grown in a Te flux. To avoid contamination, the mixing and weighting were carried out in Ar-filled glove box. Amounts of 0.5 g of Mo powder and 10 g Te were mixed and placed in an evacuated quartz ampule. The ampule was placed in a box furnace and slowly heated to 1000 °C and cooled down slowly to 800 °C followed by a hot centrifuge to remove the excess Te-flux. Single crystals were grown having a needle-like shape with a layered morphology. The as grown crystals were characterized by SEM in EDX mode for compositional analysis and by x-ray diffraction for structural analysis.

**Point contact spectroscopy.** The PCs were established at helium temperature using home-made insert by touching of thin Cu or Ag wire to a cleaved surface of MoTe$_2$ flake by a scalpel at room temperature or contacting by wire edge of plate-like samples (see Figure 8(a)). Thus, we have measured "hard" contacts between normal metal and MoTe$_2$. Additionally, we fabricated homo-contacts contacting two pieces of MoTe$_2$. Likewise, so called "soft" PCs were prepared by putting a tiny drop of silver paint on the freshly cleaved surfaces of MoTe$_2$ (see Figure 8(b)) at room temperature. We investigated in total 36 PCs, among them were 20 "hard" contacts MoTe$_2$–Ag (or Cu), 10 "soft" PCs and 6 homo-contacts MoTe$_2$–MoTe$_2$.

We measured the current-voltage (*I–V*) characteristics of PCs and their first $dV/dI$ and second $d^2V/dI^2$ derivatives in the temperature range between 2 and 10 K in magnetic field up to a few Tesla. The $dV/dI(V)\equiv R(V)$ or differential resistance and the $d^2V/dI^2(V)$ or PC spectrum were recorded by sweeping the dc current *I* on which a small ac current *i* was superimposed using facilities with set of multimeters, lock-in amplifiers, signal generator etc. An example of the electronic circuits for measuring the PC characteristics is shown in Fig. 4.3 of Ref. [13].

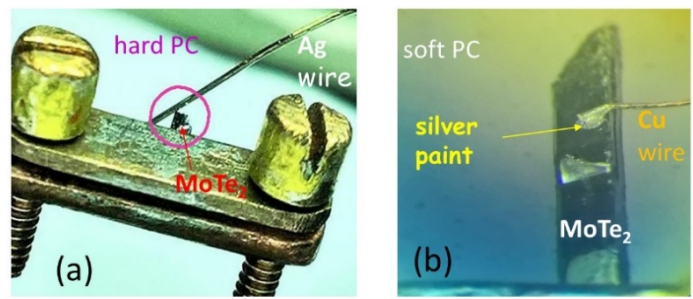

**Figure 8**. (a) Preparation of "hard" PC. The MoTe$_2$ flake is clamped between two copper plates and is touched by thin 0.1 mm Ag wire. (b) Silver paint drop is putting on cleaved MoTe$_2$ and it is wired by thin 0.07 mm Cu wire.

## Acknowledgments

Yu.G.N., O.E.K., D.L.B., S.A., I.M., I.C, S.-L.D. and D.V.E. acknowledge funding by Volkswagen Foundation. Yu.G.N., O.E.K., and D.L.B. are grateful for support by the National Academy of Sciences of Ukraine under project Ф4-19 and would like to thank the IFW-Dresden for hospitality and K. Nenkov for technical assistance.

# References

[1] Wilson J A and Yoffe A D 1969 The Transition Metal Dichalcogenides. Discussion and interpretation of the observed optical, electrical and structural properties *Adv. Phys.* **18** 193.

[2] Ali M N *et al* 2014 Large, non-saturating magnetoresistance in $WTe_2$ *Nature* **514** 205.

[3] Qi Y *et al* 2016 Superconductivity in Weyl semimetal candidate $MoTe_2$ *Nat. Commun.* **7**, 11038.

[4] Pan X-C *et al.* 2015 Pressure-driven dome-shaped superconductivity and electronic structural evolution in tungsten ditelluride *Nat. Commun.* **6** 7805.

[5] Soluyanov A *et al* 2015 Type-II Weyl semimetals *Nature,* **527**, 495.

[6] Yan B and Felser C 2017 Topological Materials: Weyl Semimetals *Annu. Rev. Condens. Matter Phys.* **8**, 337.

[7] Zandt T, Dwelk H, Janowitz C and Manzke R 2007 Quadratic temperature dependence up to 50K of the resistivity of metallic $MoTe_2$ *J. Alloys Compd.* **442**, 216.

[8] Sun Y, Wu S-C, Ali M N, Felser C. and Yan B 2015 Prediction of the Weyl semimetal in the orthorhombic $MoTe_2$ *Phys. Rev. B* **92**, 161107.

[9] Berger A N *et al.* 2018 Temperature-driven topological transition in 1T'-$MoTe_2$ *npj Q Mater.* **3** 2.

[10] Lin C-L *et al* 2018 Quasiparticle Scattering in Type-II Weyl semimetal $MoTe_2$ *J. Phys.: Condens. Matter* **30** 105703

[11] Cho G Y, Bardarson J H, Lu Y-M and Moore J E 2012 Superconductivity of doped Weyl semimetals: Finite-momentum pairing and electronic analog of the $^3$He-A phase *Phys. Rev. B* **86**, 214514.

[12] Hosur P, Dai X, Fang Z and Qi X-L 2014 Time-reversal-invariant topological superconductivity in doped Weyl semimetals *Phys. Rev. B* **90**, 045130.

[13] Daghero D and Gonnelli R S 2010 Probing multiband superconductivity by point-contact spectroscopy *Supercond. Sci. Technol.* **23,** 043001.

[14] Blonder G E, Tinkham M and Klapwjik T M 1982 Transition from metallic to tunneling regimes in superconducting microconstrictions: Excess current, charge imbalance, and supercurrent conversion *Phys. Rev. B* **25**, 4515 (1982).

[15] Naidyuk Yu G and Yanson I K 2005 Point-Contact Spectroscopy, *Springer Series in Solid-State Sciences* Vol. 145, Springer Science-Business Media, Inc., New York.

[16] Naidyuk Yu G and Gloos K 2018 Anatomy of point-contact Andreev reflection spectroscopy from the experimental point of view (Review) *Fiz. Nizk. Temp.* **44**, 343 (2018) [Low Temp. Phys., **44**, 257 (2018)].

[17] Sheet G, Mukhopadhyay S and Raychaudhuri P 2004 Role of critical current on the point-contact Andreev reflection spectra between a normal metal and a superconductor *Phys. Rev. B* **69**, 134507.

[18] Naidyuk Yu G *et al* 2017 Superconducting gaps in FeSe studied by soft point-contact Andreev reflection spectroscopy *Phys. Rev. B* **96**, 094517.

[19] Maki K 1991 Quantum oscillation in vortex states of type-II superconductors *Phys. Rev. B* **44**, 2861.

[20] Rhodes D *et al* 2017 Bulk Fermi surface of the Weyl type-II semimetallic candidate $\gamma$-MoTe$_2$ *Phys. Rev. B* **96**, 165134 (2017).

[21] Maki K 1998 Introduction to d-wave superconductivity *AIP Conference Proceedings,* **438**, 83.

[22] Naidyuk Yu G, Yanson I K, Lysykh A A and Shklarevskii O I 1982 The electron-phonon interaction in microcontacts of gold and silver *Fiz. Nizk. Temp*. **8**, 922 [*Sov. J. Low Temp. Phys.* **8**, 464].

[23] Rybalchenko L F, Yanson I K, Bobrov N L and Fisun V V 1981 Microcontact spectroscopy of tantalum, molybdenum and tungsten *Fiz. Nizk. Temp*. **7**, 169 [*Sov. J. Low Temp. Phys*. **7**, 82].

[24] Tse G and Yu D 2017 The first principle study: structural, electronic, optical, phonon and elastic properties in bulk and monolayer Molybdenum ditelluride *J. of Nanoelectronics and Optoelectronics*, **12**, 89.

[25] Geballe T H, Matthias B T, Corenzwit E and Hull Jr G W 1962 Superconductivity in Molybdenum, *Phys. Rev. Lett.* **8**, 313.

[26] Guguchia Z *et al.* 2017 Signatures of the topological s+/− superconducting order parameter in the type-II Weyl semimetal T$_d$-MoTe$_2$ *Nat. Commun.* **8,** 1082.

[27] Jiang J *et al.* 2017 Signature of type-II Weyl semimetal phase in MoTe$_2$ *Nat. Commun.* **8**, 13973.

[28] Aggarwal L *et al.* 2017 Mesoscopic superconductivity and high spin polarization coexisting at metallic point contacts on Weyl semimetal TaAs *Nat. Comm.* **8**, 13974.

[29] Wang H *et al.* 2017 Discovery of tip induced unconventional superconductivity on Weyl semimetal *Science Bulletin* **62**, 425.

[30] Takahashi H et al. 2017 Anticorrelation between polar lattice instability and superconductivity in the Weyl semimetal candidate MoTe$_2$ *Phys. Rev. B* **95**, 100501(R).